\documentclass[12pt,reqno]{amsart}
\usepackage{amsmath,amsthm,amscd,amsfonts}

\newcounter{remark}
\newcommand{\remark}{
{\noindent\bf Remark\,\,\arabic{remark}.
\addtocounter{remark}{1}\hspace{0pt}}}

\newfont{\eu}{eusb10 at 12pt}

\newcommand{\omi}[2]{\omega_{#1#2}}
\newcommand{\oms}[2]{\omega^{#1#2}}

\newcommand{\gam}[3]{\gamma^#1_{#2#3}}

\newcommand{\diff}[1]{Diff{}_{\mathrm{loc.}}^{\mathrm{#1}}}
\newcommand{\e}{\mbox{e}}
\newcommand{\Rset}{\mathbb{R}}
\newcommand{\sg}{{\sigma}}
\newcommand{\GammaG}{\Gamma_{\mathrm{\widehat{G}}}}
\newcommand{\tsm}{{\displaystyle{{T^\ast}\!\!{\mathcal{M}}}}}
\newcommand{\m}{{\mathcal{M}}}
\newcommand{\Su}{{\mathcal{S}^1_c}}

\newcommand{\A}{{\mathcal{A}}}
\newcommand{\B}{{\mathcal{B}}}
\newcommand{\F}{{\mathcal{F}}}
\newcommand{\G}{{\mathcal{G}}}
\newcommand{\I}{{\mathcal{I}}}
\newcommand{\J}{{\mathcal{J}}}

\newcommand{\Pol}{{\mathcal{P}}}
\newcommand{\Lag}{{\mathcal{L}}}
\newcommand{\Sz}{{\mathcal{S}^0_c}}
\newcommand{\tsms}{\underline{T^\ast\!\!\m}}
\newcommand{\Oc}{\widehat{\Omega}}
\newcommand{\Tc}{\widehat{\Theta}}
\newcommand{\kc}{\hat{\kappa}}
\newcommand{\bfc}{{\boldsymbol{c}}}
\newcommand{\bfah}{{\boldsymbol{\hat{a}}}}
\newcommand{\bfa}{{\boldsymbol{a}}}
\newcommand{\bftauh}{\hat{\boldsymbol{\tau}}}
\newtheorem{theorem}{Theorem}
\newtheorem{definition}{Definition}

\begin{document}
\title{Pfaff systems theory and the unifications of gravitation and
electromagnetism}
\author{Jacques L. Rubin} 
\address{Institut du
Non-Lin\'eaire de Nice (INLN), UMR 129
C.N.R.S.-Universit\'e de Nice-Sophia-Antipolis, 1361 route
des Lucioles, 06560 Valbonne, France.} 
\email{rubin@inln.cnrs.fr}
\begin{abstract} 
We show in the framework of Pfaff systems
theory, the functional dependences of the general analytic
solutions of a suitable system of involutive differential equations describing the
differences between  the analytic solutions of the conformal and ``Poincar{\'e}" Lie
equations. Then we ascribe to the infinitesimal variations of the parametrizing functionals
some physical meanings as the electromagnetic and gravitation potentials. We also deduce
their corresponding fields of interactions together with the
differential equations they must satisfy. Then we discuss on
various possible physical interpretations.
\end{abstract} 
\maketitle
\section{Introduction}  
In this paper,  we present  results about smooth deformations by
parametrizing functionals of the general solutions of
the conformal Lie equations and propose a model as well
as suggestions for a unification of electromagnetism and
gravitation. This unification also has  its roots, first
in  the conformal Lie structure that has extensively
been studied first, by J. Gasqui \cite{gasqui} and J.
Gasqui \& H. Goldschmidt \cite{gasquigold}, and second,
in the non-linear cohomology of Lie equations studied
by   H. Goldschmidt \& D. Spencer \cite[see references
therein]{gold}. Meanwhile we only partially refer to
some of its aspects since it concerns meanly the general
theory of Lie equations developed by B. Malgrange
\cite{malgrange}, A. Kumpera \& D. Spencer
\cite{kumperaspencer} and not exactly the set of PDE's
we present. Indeed these latter are not the conformal
Lie equations themselves but a kind of ``residue" coming from
the comparison with the ``Poincar\'e Lie equations". Consequently
we use the framework of Pfaff systems theory rather
than the Spencer theory of Lie equations. The results we
give in this paper, can be obtained in an equivalent way
by using the Spencer theory but in a very cumbersome way
\cite{rubin}. The Pfaff systems theory is  lighter and
is used in the context of solutions of PDE's  given by
formal series.  It is from our opinion completely
equivalent but much more simple. In fact we think that
the procedure  we develop is similar to the one
presented  by  I. G. Lisle, G. J. Reid \& A. Boulton
\cite{lisle}. In fact,  it is a Spencer theory but without its complex and
boring  terminology and concepts, together with the inherent difficulties 
coming when applied. Furthermore, this work is the result
of informal reflections about an increasing amount of
contradictions and incoherencies  concerning mainly the
concept of relativistic interaction, (that we find more
and more serious) in the field of quantum physics as well
as in classical physics. We refer the reader to a
description of these contradictions in relations to F.
Lur\c cat's \cite{lurcat}, J.-M. L\'evy-Leblond's works
\cite{levyleblond} in chapter 2. The latter 
gives also the main goals of our approach and our
initial motivations in the field of solid state physics
and in particular in the anyons theory in high-$T_c$
superconductors. It can be over-reading at a first
glance, although we give a particular spotlight on links
between unifications of forces and physics of crystals,
and we think it is unusual.\par In the chapter 3, we
present well-known results of the conformal  Lie
structures that are necessary for our purpose. Therein we
give the set of systems of PDE's from which we start our
study. It can be again over-reading by experts of the
conformal geometries, since we essentially give the origins
of the various PDE's we consider. They were obtained and detailed in H. Weyl
\cite{weyl}, K. Yano
\cite{yano} and J. Gasqui \& H. Goldschmidt works
\cite{gasquigold}. No new results are given concerning
the conformal and Poincar{\'e}  Lie structures. In fact,
in this presentation, our goal is meanly to highlight the
relations between these two kind of structures, as well
as their links with physics of gravitation.\par Chapter
4, the core of this paper, begins with the formal series
and the Pfaff systems of 1-forms we need and from which
the functional dependences of the general analytic
solutions appear, as well as their physical meanings. This
formalism can be viewed as as  kind of Spencer theory or
a differential rings theory. Then we can build up 
 differential sequences similar to the Spencer
ones, but well-known in Pfaff systems theory in contrary
to the latter.\par Chapter 5 deals with the construction
of the differential sequences. Then in the last chapters we
conclude  with suggestions for applications of these
results to unifications of interactions and cosmologic
considerations.
\section{Goals and problems: The physics of crystals and a
relativistic  phenomenology of anyons}
Our initial motivation shall be seen as extremely far
from the problems with unifications. Actually, we were more concerned
in a simple minor model of a relativistic phenomenology of creation
of anyons, accurate for certain crystals \cite{rub94}.  At the
origin of this process of creation, we suggested the
kinetico-magnetoelectrical effect as described by  E.
Asher \cite{asher} and which has its roots in the former
Minkowski works about the relations between tensors of
polarization $\Pol$ and Faraday tensors $\F$ in a moving
material of optical index 
$n\neq1$. These relations can be established by 
turning the following
diagram  into a commutative one:\vfill\eject
\[
\begin{CD}
\F'@>\Lambda>>\F\\
@V\Upsilon'VV@VV\Upsilon V\\
\Pol'@>>\Lambda>\Pol
\end{CD}
\]
where $\Lambda$ is a Lorentz transformation, allowing us
to shift from a frame 
$R'$ to a frame $R$, and $\Upsilon'$ and $\Upsilon$ are
respectively  the tensors of susceptibility within those
two frames, as well supposing $\Pol$ (or $\Pol'$) 
linearly depending on $\F$ (respectively
$\F'$). Resulting from this commutativity, the tensor
$\Upsilon$ linearly depends on $\Upsilon'$ in
general and also on a velocity 4-vector $U$ associated to
$\Lambda$ ({\it e.g.} the relative velocity 4-vector
between
$R$ and
$R'$). In assimilating $R'$ to the moving crystal frame
and $R$ to the laboratory frame, then to an applied
electromagnetic field $\F$ fixed in $R$, corresponds in
$R'$ a field of polarization $\Pol$ which varies in
relation to $U$. This is the so-called
{\it ``kinetico-magnetoelectrical effect"}. \par Parallel to this
phenomenon,  A. Janner \& E. Asher  studied the concept
of relativistic point symmetry in polarized crystals
\cite{jannerasher}. Such a symmetry is defined, on the
one hand, by a given discrete group $G$, sub-group of the
so-called Shubnikov group $O(3)1'$ (where $1'$ denotes the time inversion)
associated with the crystal, and on the other hand, as satisfying the
following properties: to make this relativistic symmetry
exists, there must be a $H(\Pol)$ non-trivial group of
Lorentz transformations  depending on $\Pol$, for which $G$
is a normal  sub-group, and that leaves the tensor of
polarization $\Pol$ invariant. In other words, if
$N(G)$ is the normalizer of $G$ in the Lorentz group $O(1,3)$, and
$K(\Pol)$ the sub-group of $O(1,3)$ leaving $\Pol$
invariant, then $H(\Pol)$ is the maximal sub-group such
that:
\begin{align*}
&H(\Pol)\subseteq K(\Pol)\cap N(G)\,,\\
&H(\Pol)\cap O(3)1'=G\,.
\end{align*}
We can prove that $H(\Pol)$ is about to exist only if a
particular non-vanishing set $V$ of velocity 4-vectors,
invariant by the action of
$G$, is present and consequently compatible with a
kinetico-magnetoelectrical effect \cite{asher}. Therefore, if there
is an interaction between moving particles in the crystal and the
polarization $\Pol$, then the trajectories and $\Pol$ are
obviously modified, and so is $H(\Pol)$. In this process,
only the group $N(G)$ is conserved so that the
polarization and the trajectories are deducible  during
the time by the action of $N(G)$.\par  As we shall
stipulate later on, the existence of an interaction will
emerge due to a correlation between the  position
3-vectors $\vec r$ of the charge carriers and a
particular 3-vector $\vec w$ ($\notin V$ in general)
associated with $\Pol$; $\vec w$ becoming then a function
of $\vec r$. In order to allow a cyclotron-type motion
which is implicit within the theory of anyons, the group
$N(G)$ must contain the group $SO(2)$ and the latter must
also non-trivially act on all the groups $H(\Pol)$
associated to
$G$. Then, only 12 groups $G$ are compatible with such a
description \cite[see the table therein and the 12 groups
for which toroidal phases exist]{rub94,rub93}:
\[
1,\ 2',\ m,\ m',\ {\bar 1}',\ 2'/m,\  {\bar 3}',\
2/m',\ 4/m',\ 6/m',\ {\bar 4}',\ {\bar 6}'\,.
\] 
It has to be noticed that among these symmetries allowing
toroidal moments in crystals, most of them impose
coupling of pairs of toroidal moments, each one
associated to a carrier, to avoid a non-vanishing total
orbital moment not compatible with the symmetry. 
Then, a coupling of charge carriers can
occur without the need of any kind of particular (always
unknown) interactions ! It can be viewed as an
alternative  to BCS type couplings. We can speak about a kind of{\it
``inverse kineto-magnetoelectric effect"} since the crystals are
motionless in contrary to the carriers.\par In fact, throughout this
development, we implicitly use a principle of equivalence similar to the one
formulated in general relativity: one cannot distinguish a cyclotron-type
motion in a constant polarization field from a uniform rectilinear motion in
a field of polarization varying in time by action of the normalizer
$N(G)$. From an other point of view,
the interaction is considered to allow the extension of
an invariance with respect to $H(\Pol)$ to an invariance
with respect to $N(G)$. The lack of interaction is then
what breaks down the symmetry !\par This type of reasoning
concerns in fact a large amount of physical phenomena
such as the spin-orbit interaction for instance. In this
context, the cyclotron-type motion of electrons in
anyonic states would be similar to the Thomas or Larmor
precessions (see also the Coriolis or Einstein-Bass
effects). More precisely, taking up again a computation,
analogous to the Thomas precession one \cite{bacry}
({\it e.g.}  considering as a constant the scalar product
of two tangent vectors being two parallel transports along
the trajectory
\cite{dieudon}), concerning a charge carrier at $\vec r$ with the
velocity  4-vector $U$  together in $R$, ``polarized'' by
${\vec w}({\vec r})$ such as for example  (${\xi}=(0,{\vec
\xi})_R$ constant and ${\vec \xi}\in V$):
\[
{W}=(0,{\vec w})_R\equiv 
-P\,.\,{\xi}\; \text{or}\; {}^{\ast}P\,.\,{\xi}\,,
\]
where $\Pol$ depends on $\vec r$, one can prove from
$\Pol\,.\,\xi\,.\,U\equiv W\,.\,U=cst$ ($t$ being the
laboratory frame time and ($\tilde{r}=(t,\vec{r})_R$) that:
\begin{equation}
{\frac{\,dU}{dt}}= (-e/m)F_{eff.}(\tilde{r})\,.\,U\,,
\label{i}
\end{equation}
where $m$ and $e$ are respectively the mass and the
electric charge of the carrier and
$F_{eff.}(\tilde{r})\equiv ({\vec E}_{eff.}
(\tilde{r}),{\vec B}_{eff.}(\tilde{r}))$ is an effective
Faraday tensor such that ($\gamma = (1-{\vec
w}^2)^{-{1/2}}$ and
${\vec j} =e\,d{\vec r}/dt$):
\begin{align*}
&{\vec B}_{eff.}(\tilde{r})=(m/e)
\left(\frac{\gamma}{1+\gamma}\right){\vec w}
\wedge\frac{d{\vec w}}{dt}
\equiv(m/e^2)
\left(\frac{\gamma}{1+\gamma}\right){\vec w}
\wedge\left[{\vec j}.{\vec \nabla}\right]{\vec w}\,,\\
&{\vec E}_{eff.}(\tilde{r})={\vec 0}\,.
\end{align*}
In fact we have just taken the general formula for
the Thomas precession, and then substituting the velocity
3-vector by $\vec w$ and the acceleration 3-vector by the
time derivative of $\vec w$ depending on the space
position 3-vector $\vec r$. Clearly,
$F_{eff.}$  is an element of the Lie algebra of the group
$SO(2)$ included in $N(G)$ and with
${\vec B}_{eff.}\in V$. Therefore this magnetic
field ${\vec B}_{eff.}$
 or $\vec \xi$ (up to a constant) might be
considered as the effective magnetic field of the
flux-tube $V$ generating the so-called Aharanov-B\"ohm
effect at the origin of the statistical parameter in the
anyons theory
\cite{Wilczek}. Let
us add that in general the divergence 
$\text{Div}({\vec B}_{eff.})\not =0$ so that one gets a
non-vanishing density of effective magnetic monopoles
generated by the local variations (due to the
interaction) of the polarization vector field ${\vec w}$
in the crystal. Thus an anyon would be an effective
magnetic monopole associated with a charge  carrier,
namely a dyon. Moreover, because this effective Faraday
tensor is no more a closed two-from, a non-vanishing
Chern-Simon has to be taken into account in a Lagrangian
description of anyons, from which non-vanishing
spontaneous constant currents can occur.  On that
subject, one can notice that the equation (\ref{i}) can
be rewritten  in an orthonormal system of local
coordinates  ($i,j,k= 1, ..., n$; the
$\Gamma$'s being  skew-symmetric connection symbols):
\begin{equation}
{{\dot U}^i}+
{\Gamma{}_{j,k}^{i}}\,{U^j}\,{U^k}=0
\,.
\label{torsion}
\end{equation}
We recognize the equations of the geodesics associated to a
Riemannian connection with torsion which is thus
associated to a {\it ``generalized Thomas precession"} phenomena. This
would suggest a unification in reference to the Einstein-Cartan theory but
as we shall see it is not  the case despite the appearances.
Then if we keep on with the assumption that one has to
add to the electromagnetic field a gravitational field
and that the derivatives of the fields are functions of
the fields themselves (as with the Bianchi identities
according to the non-abelian theory for example), that
means we make the assumption of the existence of a
differential sequence. In electromagnetism, it is a
matter of the de Rham sequence but  gravitation does not
interfere. The sequence integrating the latter - and
being the purpose of this paper -  might be a certain
generalizing complex like the Spencer one.
\section{The conformal finite
Lie equations}
First of all, let us assume that the group of relativity is not the
Poincar\'{e} group anymore but the conformal Lie group (we know
from Bateman and Cunningham studies \cite{bateman} that	 it is
the group of invariance of the Maxwell equations). In particular,
this involves that no changes occur shifting from a given frame
to a uniformly accelerated relative one. From a historical point
of view, that happened to be the starting point of the Weyl
theory which was finally in contradiction with experimental
data.\par Let us first call $\mathcal{M}$, the base space (or
space-time), assumed to be of class $C^\infty$, of dimension
$n\geq 4$, connected, paracompact, without boundaries, oriented and
endowed with a metric 2-form $\omega$, symmetric, at
least of class $C^2$ on
$\mathcal{M}$ and non-degenerated but not necessarily
definite positive.  We also  assume $\mathcal{M}$ to have
a constant Riemannian scalar curvature. From these
considerations,
$\mathcal{M}$ has a pseudo-Riemannian or Riemannian
structure because of  these metric features. 
\par
The conformal finite Lie equations are deduced from the conformal
action on the me\-tric defining a  pseudo-Riemannian
structure on $\mathcal{M}$. Let us consider ${\hat
f}\in\diff{1}({\mathcal{M}})$, the set of local diffeomorphisms of
$\mathcal{M}$ of class $C^1$, and any function
$\alpha\in{C^0}({\mathcal{M}},{\Rset})$. Then if ${\hat
f}\in{\GammaG}$ ($\GammaG$ being the pseudogroup of local
conformal bidifferential maps on ${\mathcal{M}}$), ${\hat f}$ is
a solution of the following  system of PDE's (in fact other PDE's must be
satisfied to completely define $\GammaG$ as one shall see in the
sequel):
\begin{equation} 
{{\hat f}^\ast}\omega=
\e^{2\alpha}\omega
\label{1}
\end{equation}
with $\det(J({\hat f}))\not=0$, and where $J({\hat f})$ is
the Jacobian of $\hat f$, and ${\hat f}^{\ast}$ is the
pull-back of $\hat f$. Also only the $\e^{2\alpha}$
positive functions are  taken into account because of the
previous assumption that only one orientation is chosen
and kept on
$\mathcal{M}$, therefore we consider only the $\hat f$'s keeping
the orientation.  We  have to recall that $\alpha$ is a
varying function depending on each $\hat{f}$. We denote
$\tilde{\omega}$ the metric on ${\mathcal{M}}$ such as by definition:
$\tilde{\omega}\equiv\e^{2\alpha}\omega$, and we agree to put a
tilde on each tensor or geometrical ``object" relative to or
deduced from this  metric $\tilde{\omega}$. Let us notice that
the  latter depends on a fixed given element ${{\hat
f}}\in{\GammaG}$ when expressed without the scalar function
~$\alpha$.
\par Now, doing a first prolongation of the system (\ref{1}), we
deduce other second order PDE's connecting the covariant
derivations of Levi-Civita $\nabla$ and
$\widetilde{\nabla}$ respectively associated to $\omega$
and $\tilde{\omega}$. These new  differential equations are 
\cite{gasqui} $\forall\, X,\,Y\in
T{\mathcal{M}}$, 
\begin{equation}
{{\widetilde{\nabla}}_X}Y={\nabla_X}Y+d\alpha(X)Y+d\alpha(Y)X-
\omega(X,Y)\,{{}_\ast}d\alpha, \label{3} \end{equation}
where $d$ is the exterior differential and ${{}_\ast}d\alpha$
is the dual vector field of the 1-form $d\alpha$ with respect
to the metric $\omega$, {\it e.g.} such as $\forall\,
X\in T{\mathcal{M}}$,
\begin{equation}
\omega(X,{{}_\ast}d\alpha)=d\alpha(X)\,.
\label{4} \end{equation}
Prolonging again and using the definition of the Riemann tensor
$\rho$ associated to $\omega$, one obtains the following relation
$\forall\,X,Y\in{C^1}(T{\mathcal{M}})$, 
$\forall\,Z\in{C^2}(T{\mathcal{M}})$,
$\omega\in{C^2}(S^2 T^{\ast}{\mathcal{M}})$,
$\forall\,\alpha\in{C^2}({\mathcal{M}},{\Rset})$ and
$\forall\,{\hat f}\in~\diff{3}({\mathcal{M}})$, 
\begin{multline} 
\tilde{\rho}(X,Y).Z=
\rho(X,Y).Z+\omega(X,Z){\nabla_Y}({{}_\ast}d\alpha)+\\
\left\{\omega({\nabla_X}({{}_\ast}d\alpha),Z)+
\omega(X,Z)d\alpha({{}_\ast}d\alpha) \right\}Y-\\ 
\left\{\omega({\nabla_Y}({{}_\ast}d\alpha),Z)+
\omega(Y,Z)d\alpha({{}_\ast}d\alpha) \right\}X+\\
\left\{d\alpha(X)\omega(Y,Z)- d\alpha(Y)\omega(X,Z)
\right\}{{}_\ast}d\alpha+\\
\left\{d\alpha(Y)X-d\alpha(X)Y\right\}d\alpha(Z)-
\omega(Y,Z){\nabla_X}({{}_\ast}d\alpha). 
\label{5} 
\end{multline}
Then we Consider ${\mathcal{M}}$ to be conformally flat because
of  the constant Riemann scalar curvature, the Weyl tensor
$\tau$ associated with $\omega$ vanishes. Hence, the Riemann
tensor $\rho$ can be rewritten
$\forall\,U\in{C^0}(T{\mathcal{M}})$,
$\forall\,X,Y\in{C^1}(T{\mathcal{M}})$,
$\forall\,Z\in{C^2}(T{\mathcal{M}})$ as:
\begin{multline} 
\omega(U,\rho(X,Y).Z)={\frac{1}{(n-2)}}
\left\{\omega(X,U)\sigma(Y,Z)-
\omega(Y,U)\sigma(X,Z)\right.+\\
\left.\omega(Y,Z)\sigma(X,U)-
\omega(X,Z)\sigma(Y,U)\right\}, 
\label{6} 
\end{multline}
where $\sigma$ is what we call the ``Yano tensor" (see the tensor
``L" in \cite{yano} differing from $\sigma$ by the fraction
$1/(n-2)$, or the tensor $\omega$ in \cite{gasquigold} formula
(3.12) p.68) defined
$\forall\,X,Y\in{C^1}(T{\mathcal{M}})$ by
\begin{equation}
\sigma(X,Y)={\rho_{\mathrm{ic}}}(X,Y)-
{\frac{\rho_{\mathrm{s}}}{2(n-1)}}\omega(X,Y),
\label{7} 
\end{equation}
where $\rho_{\mathrm{ic}}$ is the Ricci tensor and $\rho_{\mathrm{s}}$
is the Riemann scalar curvature. Consequently, the system of PDE
(\ref{5}) can be rewritten as a first order system of PDE
concerning $\sigma$. In order to do this, we first define
two suitable trace operators, used in the sequel to
obtain the
$\tilde{\rho}_{\mathrm{ic}}$ and $\tilde{\rho}_{\mathrm{s}}$
tensors and finally the $\tilde{\sigma}$ tensor. Let us denote
$\mathrm{Tr}^1$ the trace operator defined such that for any
vector bundle $E$ over ${\mathcal{M}}$ we have: 
\[
{\mathrm{Tr}^1}:T{\mathcal{M}}\otimes\tsm\otimes{E}
\longrightarrow{E}
\] with
\,${\mathrm{Tr}^1}(X\otimes\alpha\otimes\mu)=\alpha(X)\mu$\, for
any $X\in T{\mathcal{M}}$, $\alpha\in\tsm$ and $\mu\in E$. Then,
the second trace operator is the natural trace
$\mathrm{Tr}_\omega$ associated to $\omega$ and defined by:
\[{\mathrm{Tr}_\omega}:{\buildrel{2}\over\otimes}\tsm\longrightarrow
{\Rset},\]
such that \,${\mathrm{Tr}_\omega}(u\otimes v)=v({{}_\ast}u)$. Finally,
with ${\mathrm{Tr}^1}\tilde{\rho}={\tilde{\rho}_{\mathrm{ic}}}$
and ${\mathrm{Tr}_\omega}{\tilde{\rho}_{\mathrm{ic}}}=
{\tilde{\rho}_{\mathrm{s}}}$,
we deduce first from the relations (\ref{5}) and (\ref{7})
$\forall\,{\hat f}\in \diff{3}({\mathcal{M}})$, $\forall\,X,Y\in
{C^1}(T{\mathcal{M}})$ and
$\forall\,\alpha\in{C^2}({\mathcal{M}},{\Rset})$,
\begin{multline} 
\tilde{\sigma}(X,Y)=\sigma(X,Y)+\\
(n-2)\Big(d\alpha(X)d\alpha(Y)
-{\frac{1}{2}}\omega(X,Y)d\alpha({{}_\ast}
d\alpha)-\omega({\nabla_X}({{}_\ast}d\alpha),Y)\Big). 
\label{8} 
\end{multline}
In fact this expression can be symmetrized and using the torsion
free property of the Levi-Civita covariant derivations, one obtains
$\forall\,{\hat f}\in\diff{3}({\mathcal{M}})$,
$\forall\,X,Y\in{C^1}(T{\mathcal{M}})$ and
$\forall\,\alpha\in{C^2}({\mathcal{M}},{\Rset})$:
\begin{equation} 
\tilde{\sigma}(X,Y)=\sigma(X,Y)+(n-2)\Big(d\alpha(X)
d\alpha(Y)-\mu(X,Y)-{\frac{1}{2}}\omega(X,Y)
d\alpha({{}_\ast}
d\alpha)\Big),
\label{10} 
\end{equation}
into which we define $\mu\in{C^0}({S^2}\tsm)$ by:
\begin{equation}
\mu(X,Y)={\frac{1}{2}}\left[X.\,d\alpha(Y)+Y.\,d\alpha(X) -
d\alpha({\nabla_X}Y+{\nabla_Y}X)\right]. \label{9}
\end{equation}
From the proposition 5.1 due to J. Gasqui and H.
Goldschmidt \cite{gasquigold} and  the well-known theorem
of H. Weyl on equivalence of conformal structures
\cite{weyl}, and because of the Weyl tensor vanishing, the
differential equation (\ref{1}) is formally integrable.
Also, since the  Riemann scalar curvature is assumed to be
a constant, the Yano tensor satisfies the following
relation (see for instance, formula (16.3) p.183 in
\cite{gasquigold}):
\begin{equation} \sigma={k_0}{\frac{(n-2)}{2}}\omega\,, \label{11}
\end{equation}
with ${\rho_{\mathrm{s}}}={k_0}\,({k_0}\in {\Rset})$. Then,
considering the system (\ref{1}), the system (\ref{10}) reduces
to  a second order system of PDE's such as
$\forall\,\alpha\in{C^2}({\mathcal{M}},{\Rset})$ and
$\forall\,X,Y\in{C^1}(T{\mathcal{M}})$ we have:
\begin{equation} 
\mu(X,Y)={\frac{1}{2}}\left\{
\left[{k_0}\left(1-{\e^{2\alpha}}\right)
-d\alpha({{}_\ast}d\alpha)
\right]\omega(X,Y)\right\}
+d\alpha(X)d\alpha(Y)\,. 
\label{12} 
\end{equation}
Thus, we have  series of PDE's deduced from (\ref{1}). In
particular the system made of the PDE's (\ref{1}) and
(\ref{3}) is formally integrable and even involutive of finite
type because the symbol ${\widehat{M}}_3$ of order 3 is vanishing
(see $\mbox{g}^c_3$ in \cite{gasquigold}). But there are
alternative versions of these PDE's in which the function
$\alpha\in{C^2}({\mathcal{M}},{\Rset})$ doesn't appear. These latter are the following: from the system
(\ref{1}), one deduces, $\forall\,{\hat
f}\in\diff{1}({\mathcal{M}})$: 
\begin{equation} 
{{\hat f}^\ast}{\hat
\omega}/(\det(J({\hat
f})))^{2/n}={\hat \omega}\,, 
\label{14} 
\end{equation}
with $\det(J({\hat f}))\not= 0$ and
${\hat\omega}=\omega/{{\det(\omega)}^{1/n}}$, and by
setting:
\[
{{}^{\hat f}\!B}(X,Y)\equiv{{\widetilde{\nabla}}_X}Y-{\nabla_X}Y\,,
\]
where ${{}^{\hat
f}\!B}\in{T{{\mathcal{M}}}}\otimes{S^{2}}{T^\ast}\!\!{\mathcal{M}}$
is the second fundamental quadratic form associated to
$\hat f$,  one obtains
\cite{gasqui} from (\ref{3}) 
$\forall\,X,Y\in{C^1}(T{\mathcal{M}})$,
$\forall\,{\hat f}\in\diff{3}({\mathcal{M}})$ the second order
differential equation:
\begin{equation}
n\,{{}^{\hat f}\!B}=2\,{\mathrm{Tr}^1}({{}^{\hat f}\!B}) - 
\omega\,{{}_\ast}\!{\mathrm{Tr}^1}({{}^{\hat f}\!B})\,.
\label{15}
\end{equation} 
Then  the conformal 
Lie pseudogroup $\GammaG$ with $\tau=0$ is the set of functions
${\hat{f}}\in\diff{3}({\mathcal{M}})$ satisfying the 
involutive system of PDE's (\ref{14}) and (\ref{15}). One points out again 
one has the supplementary differential equation (\ref{12}) because of the additional assumption the Riemann scalar
curvature is a constant. This differential equation is the main point allowing to
build up a relative complex of smooth deformations
associated to the unification model.\par
The  Poincar\'{e} pseudogroup
corresponds to the case for which $\alpha=0$. The symbol
$M_2$ of order 2 of this pseudogroup vanishes and therefore it is
involutive, and  the
Poincar\'{e} pseudogroup is not formally integrable unless
$\rho_{\mathrm{s}}$ is a constant \cite{gasquigold}. 
In an orthonormal system of coordinates, the
PDE's (\ref{1}), (\ref{3}) and (\ref{12}) can be
written, with
$\det(J({\hat f}))\not= 0$ and $i,j,k=1,\cdots,n$ as

\begin{subequations}
\begin{align}
&\sum_{r,s=1}^n\omega_{rs}(\hat
f)\,\,\hat{f}^{r}_{i}\hat{f}^{s}_{j}=
e^{2\alpha}\omega_{ij}\,,
\label{f1}\\ 
&\hat{f}^k_{ij}+
\sum_{r,s=1}^n{}\gamma^{k}_{rs}(
\hat{f})\,\,\hat{f}^{r}_{i}\hat{f}^{s}_{j}=
\sum_{q=1}^n{}\hat{f}_q^k\left(
\gamma^{q}_{ij}+\alpha_i\delta^q_j+
\alpha_j\delta^q_i-\omega_{ij}\alpha^q\right)\,,
\label{f2}\\
&\alpha_{ij}=\frac{1}{2}
\Big\{k_0(1-\mbox{e}^{2\alpha})-
\sum_{k=1}^n\alpha^k\alpha_k
\,\Big\}\omega_{ij}+
\alpha_i\alpha_j\,,\label{third}
\end{align}
\label{confeq}
\end{subequations}
where $\delta^i_j$ is the Kronecker tensor, and where one
denotes as usual
 $\hat{f}^i_j\equiv\partial \hat{f}^i/\partial
x^j\equiv\partial_j \hat{f}^i$, etc..., $T_k=\sum_{h=1}^n
T^h\,\omega_{hk}$ and $T^k=\sum_{h=1}^n
T_h\,\omega^{hk}$ for any tensor $T$ where $\omega^{ij}$
is the inverse metric tensor, and
$\gamma$ is the  Riemann-Christoffel form associated to
$\omega$. This is the set of our starting equations.\par
\begin{definition}
We call the \emph{``third order system"}, the system of
PDE's for $\hat{f}^i_J$ ($|J|\leq3$) only defined by the
relations of ``order 3" deduced from the first
prolongation of
\eqref{f2}. In this system, the
derivatives of $\alpha$ do not appear. These 
PDE's  are ``shaped" like ($i,j,k,h,r=1,\cdots,n$):
$\hat{f}^i_{jkh}\equiv$ polynomial of terms
$\hat{f}^r_K$ ($|K|\leq2$) with the derivatives of the
Riemann-Christoffel symbols up to order two, the metric
$\omega$, the constants $n$  and $k_0$ as coefficients.
\end{definition}
Under a change of coordinates with a conformal
application
$\hat{f}$, the function
$\alpha(\equiv\tilde{\alpha}_0)$ and the tensor
$\tilde{\alpha}_1\equiv\{\alpha_1,\cdots,\alpha_n\}$ are
transformed onto ``primed" functions and tensors such as
($j=1,\cdots,n$):
\begin{subequations}
\begin{align}
&\alpha'(\hat{f})=\alpha-\frac{1}{n}\ln{|}\det
J(\hat{f})\,{|}\,,\\
&\sum_{i=1}^{n}\hat{f}^i_j\,{\alpha'}_i(\hat{f})=\alpha_j-\sum_{k,l=1}^{n}
\frac{1}{n}\,\hat{f}^{-1\,k}_l(\hat{f})\,\hat{f}^l_{kj}\,,
\end{align}
\label{albet}
\end{subequations}
which shows essentially the affine feature of these
``geometrical objects", and in particularly that the
tensor $\tilde{\alpha}_1$ is
associated to the second order of derivation of $\hat{f}$.
Then it could be considered as an acceleration tensor. In
that case, a change of acceleration would keep conformal
physic laws invariant. It would be closed to Einsteinian
relativity.\par From a mathematical point of view, it is
important to notice that
$\mu$ or equivalently the tensor
$\tilde{\alpha}_2\equiv\{\alpha_{ij},i,j=1,\cdots,n\}$
might be considered as an Abraham-E\"{o}tv\"os type tensor
\cite{abraham}, leading to a first physical
interpretation (up to a constant for units and with
$n=4$) of
$\tilde{\alpha}_1$ as the acceleration 4-vector of gravity
and
$\alpha$ the Newtonian potential of
gravitation. On the other hand,
$\alpha$ being associated with the dilatations perhaps
it might be considered as a relative Thomson type
temperature (again up to a constant for units):
\begin{equation*} 
\alpha=\ln({T_0}/T)\,, 
\end{equation*}
where $T_0$ is a constant temperature of reference associated
with the base space-time ${\mathcal{M}}$. A first question arises
about this temperature $T_0$: can we consider
it as the 2.7 K cosmic background temperature ?  Also from
the transformation law of this function, it would involve
the temperature $T$ would not be a conformal invariant but only a
Lorentz (relativistic) invariant since in that case
$|\det J(\hat{f})|=1$, a  question  which seems always to
be under investigations for instance in ``hot QCD" theories
although low temperatures ({\it e.g.} low energies) are
considered as relativistic invariants.\par In fact, 
$\alpha$ could be considered  either as a Newtonian 
potential of gravitation or a  temperature  (if it is
physically available),   or more judiciously as a sum of
the two. It appears that physical interpretations could be
made at two levels: a ``global" one at universe scales,
and a ``local" one in considering forces of gravitation.
In the same way, a second question arises at the ``global"
level about the meaning of the tensor $\tilde{\alpha}_1$ 
up to units: since it might be an acceleration, would it
be the acceleration of the inflation process ? Then this
acceleration would describe the cosmic temperature
background evolution and perhaps might be equivalently
associated to a cosmic (repulsive ?) background of
radiation of  gravitational waves. As we shall see, the
various dynamics will  depend only on the physical
interpretations of these two parameters.
\section{The functional dependence}
Now we look  on formal series, solutions of the system of
PDE's (\ref{confeq}). We know these series will be
convergent in a suitable open subset because the system
is involutive. Nevertheless we need of course to know the
Taylor coefficients. For instance we can choose  for the
applications $\hat{f}$ and the functions
$\alpha$  the following series at a point
$x_0\in\m$:
\begin{align*}
\hat{f}^i(x):\quad&
S^i(x,x_0,\{\hat a\})=\sum_{|J|\geq0}^{+\infty}
\hat{a}^i_J(x-x_0)^J/|J|!\,,\\
\alpha(x):\quad&
s(x,x_0,\{c\})=\sum_{|K|\geq0}^{+\infty} 
c_K(x-x_0)^K/|K|!\,,
\end{align*}
with $x\in U_{x_0}\subset\m$ being a suitable open 
neighborhood of $x_0$ to insure the convergence of
the series,
$i=1,\cdots,n$, $J$ and
$K$ are multiple index notations such as $J=(j_1,\cdots,j_n)$, 
$K=(k_1,\cdots,k_n)$ with $|J|=\sum_{i=1}^n j_i$ and
similar expressions for $|K|$, $\{\hat{a}\}$ and
$\{c\}$ are the sets of Taylor coefficients and
$\hat{a}^i_J$ and $c_K$ are real values and not functions
of $x_0$, though of course there are values of functions at
$x_0$.
\subsection{The ``c" system}
We call the ``c" system, the system of PDE's
(\ref{third}). It is from this set of PDE's the potentials
and fields of interactions could occur. From the series $s$,
at zero-th order one obtains the algebraic equations
($i,j=1,\cdots,n$; $\bfc_1=\{c_1,\cdots,c_n\}$):

\begin{equation}
c_{ij}=\frac{1}{2}\Big\{k_0(1-\e^{2c_0})-
\sum_{k,h=1}^n\omega^{kh}(x_0)c_h c_k
\,\Big\}\omega_{ij}(x_0)+
c_i c_j\equiv F_{ij}(x_0,c_0,\bfc_1)\,,
\label{c2eq}
\end{equation}
and it follows the $c_K$'s such that $|K|\geq2$, will
depend recursively only on $x_0$, $c_0$ and $\bfc_1$.
It is none but the least the meaning of involution of
so-called involutive systems and this recursion property
can be also related by analogy to Painlev{\'e} tests. Hence the
series for
$\alpha$ can be written as
$s(x,x_0,c_0,\bfc_1)$. By varying $x_0$, $c_0$ and
$\bfc_1$, we can change or not the function $\alpha$.
Let $J_1$ be the 1-jets affine bundle of the
$C^\infty$ real valued functions on $\m$. Then it exists
a subset associated to
$\bfc_0^1\equiv(x_0,c_0,\bfc_1)$, we denote
$\Su(\bfc_0^1)$, of elements
$({x'}_0,{c'}_0,\bfc{'}_1)\subset J_1$, such that there
is an open neighborhood
$U(\bfc_0^1)\subset \Su(\bfc_0^1)$, projecting on $\m$ in a
neighborhood of a given
$x\in U_{x_0}$, for which for all $({x'}_0,{c'}_0,\bfc{'}_1)\in U(\bfc_0^1)$ then
$s(x,x_0,c_0,\bfc_1)=s(x,{x'}_0,{c}{'}_0,\bfc{'}_1)$.
Is this subset $\Su(\bfc_0^1)$ a submanifold of $J_1$,
involving for a fixed $x$, the variation $ds$ with respect
to $x_0$,
$c_0$ and $\bfc_1$ is vanishing ?
From $ds\equiv0$ it
follows that ($k=1,\cdots,n$):
\begin{align*}
&\sg_0\equiv dc_0-\sum_{i=1}^nc_idx_0^i=0\,,\\
&\sg_k\equiv dc_k-
\sum_{j=1}^n F_{kj}(x_0,c_0,\bfc_1)\,dx_0^j=0\,.
\end{align*}
We recognize a Pfaff system, we denote
$P_c$, generated by the 1-forms $\sg_0$ and $\sg_k$, and
the meaning of their vanishing, {\it e.g.} the solutions
$\alpha$ do not change for such variations of $c_0$,
$\bfc_1$ and $x_0$. Also, as it can be easily
verified, the Pfaff system $P_c$ is integrable since the
Fr\"obenius condition is satisfied, and all the
prolongated 1-forms $\sg_K$ ($K\geq2$) will be linear
combinaisons of these $n+1$ generating forms from the
recursion property of involution. Then the subset
$\Su(\bfc_0^1)$ of dimension $n$ containing a particular
element
$\bfc_0^1\equiv(x_0,c_0,\bfc_1)$ is a submanifold of
$J_1$ that we call by lack the ``solutions submanifold
of order one at $\bfc_0^1$". It is a particular leaf
of, at least, a local foliation on $J_1$ of codimension
$n+1$.\par From the integrability of
$P_c$, one deduces that it exists on
$J_1$, local systems of coordinates
$(x_0,\tau_0,\tau_1,\cdots,\tau_n)$ such that each leaf
$\Su(\bfc_0^1)$ is a submanifold for which $\tau_0=cst$
and $\tau_i=cst$ $(i=1,\cdots ,n)$, involving all the
series $s(x,{x'}_0,{c'}_0,\bfc{'}_1)$ with
$({x'}_0,{c'}_0,\bfc{'}_1)\in\Su(\bfc_0^1)$ equal a same
function $s'(x,\tau_0,\boldsymbol{\tau}_1)$ 
($\boldsymbol{\tau}_1\equiv\{\tau_1,\cdots,\tau_n\}$).
Then the difference
$s(x,x_0,c_0,\bfc_1)-s(x,{x'}_0,{c'}_0,\bfc{'}_1)$
satisfies the relation:
\begin{equation}
s(x,x_0,c_0,\bfc_1)-s(x,{x'}_0,{c'}_0,\bfc{'}_1)=
s'(x,\tau_0,\boldsymbol{\tau}_1)-
s'(x,{\tau'}_0,\boldsymbol{\tau}{'}_1)\,,
\label{diffs}
\end{equation}
with ($i=1,\cdots,n$)
\begin{align*}
&\triangle_0\tau\equiv{\tau'}_0-\tau_0=
\int_{\bfc^1_0}^{{\bfc'}^1_0}\!
\sg_0\,,\\
&\triangle_i\tau\equiv{\tau'}_i-\tau_i=
\int_{\bfc^1_0}^{{\bfc'}^1_0}\!
\sg_i\,.
\end{align*}
Now, we consider the
$c$'s are values of differential functions
$\rho\,$: $c_K=\rho_K(x_0)$, as expected for Taylor
coefficients. Roughly speaking, we make a pull-back
on $\m$, inducing a projection from the subbundle of
projectable elements in
$T^\ast\!J_1$ to\linebreak
$T^\ast\!\!\m\otimes_\Rset\!J_1$. Then, we set 
(with $\boldsymbol{\rho}_1\equiv\{\rho_1,\cdots,\rho_n\}$
and no changes of notations for the pull-back):
\begin{subequations}
\begin{align}
&\sg_0\equiv\sum_{i=1}^n(\partial_i\rho_0-\rho_i)\,dx_0^i\equiv
\sum_{i=1}^n\A_i\,dx_0^i\,,
\label{rhoA}\\
&\sg_i\equiv\sum_{j=1}^n(\partial_j\rho_i-
F_{ij}(x_0,\rho_0,\boldsymbol{\rho}_1))\,dx_0^j
\equiv\sum_{j=1}^n\B_{j,i}\,dx_0^j\,,
\end{align}
\label{rhoAB}
\end{subequations}
and it follows that
\begin{align*}
&\triangle_0\tau=
\int_{x_0}^{{x'}_0}
\sum_{i=1}^n\A_i\,dx^i\,,\\
&\triangle_i\tau=
\int_{x_0}^{{x'}_0}
\sum_{j=1}^n\B_{j,i}\,dx^j\,.
\end{align*}
In particularly, if $\bfc_0^1\in\Su(0)$, the ``null"
submanifold corresponding to the vanishing solution of the
third system with $\tau_0=\tau_1=\cdots=\tau_n=0$, then 
the difference (\ref{diffs}) involves that
\begin{equation*}
\alpha(x)\equiv
s(x,{x'}_0,{c'}_0,\bfc{'}_1)=
s'(x,{\tau'}_0,\boldsymbol{\tau}{'}_1)=s(x,x\equiv
{x''}_0,{c''}_0,\bfc{''}_1)\,,
\end{equation*}
with
\begin{align*}
&{\tau'}_0=
\int_{x_0}^{{x''}_0}
\sum_{i=1}^n\A_i\,dx^i\,,\\
&{\tau'}_i=
\int_{x_0}^{{x''}_0}
\sum_{j=1}^n\B_{j,i}\,dx^j\,,
\end{align*}
and $\bfc{''}_0^1\in\Su(\bfc{'}_0^1)$.
In particularly, since we can take ${x''}_0\equiv x$ then
\begin{equation}
\alpha(x)\equiv s'\Big(x,\int_{x_0}^{x}
\sum_{i=1}^n\A_i\,{dx'}^i\,,\int_{x_0}^{x}
\sum_{j=1}^n\B_{j,1}\,{dx'}^j,\cdots,
\int_{x_0}^{x}
\sum_{j=1}^n\B_{j,n}\,{dx'}^j\Big)\,,
\label{deformpar}
\end{equation}
which shows the functional dependences of the solutions of
the ``c" system with respect to the functions $\rho_0$
and $\boldsymbol{\rho}_1$, themselves associated to the
smooth infinitesimal deformations of these solutions.
These smooth deformations can also be considered as smooth
deformations from ``Poincar{\'e} solutions" of the system
(\ref{confeq}) for which $\alpha\equiv0$, to ``conformal
solutions" whatever is $\alpha$.\par Moreover the functions
$\rho$, and consequently the functions
$\rho_0$, $\A$ and
$\B$, must satisfy additional differential equations
coming from the integrability conditions of the Pfaff
system
$P_c$. More precisely, from the relations
$d\sg_0=\sum_{i=1}^n\,dx_0^i\wedge\sg_i$,
$d\sg_i=\sum_{j=1}^n\,dx_0^j\wedge\sg_{ij}$ and 
\begin{equation}
\sg_{ij}=c_i\sg_j+c_j\sg_i-\omega_{ij}\Big\{k_0\e^{2c_0}\sg_0+
\sum_{k,h=1}^n\omega^{kh}c_h\sg_k\Big\}
\equiv\vartheta_{ij}(\bfc_0^1,\sg_J;|J|\leq1)\,,
\label{sg2}
\end{equation}
one deduces
a set of algebraic relations at $x_0$:
\begin{subequations}
\begin{align}
&\I_{k,i}\equiv \partial_k\A_i-\B_{i,k}\,,\label{IAB}\\
&\I_{k,i}=\I_{i,k}\,,
\end{align}
\begin{equation}
\J_{k,j,i}\equiv
\partial_j\B_{k,i}-
\rho_i\,\B_{k,j}-
\rho_j\,\B_{k,i}+
\omega_{ij}\left\{k_0e^{2\rho_0}\A_k+
\sum_{r,s=1}^n
\omega^{rs}\,\rho_r\,\B_{k,s}\right\}\,,
\label{JAB}
\end{equation}
\begin{align}
\J_{k,j,i}=\J_{j,k,i}\,.
\end{align}
\label{ABIJeq}
\end{subequations}  
Clearly, in these relations, $(\rho_0,\boldsymbol{\rho}_1)$
appears to be a set of arbitrary functions. In considering 
$\F$ and
$\G$ as being respectively the skew-symmetric and the
symmetric parts of the tensor of components
$\partial_i\rho_j$, then one deduces, from the symmetry
properties of the latter relations, what we call {\it the
first set of differential equations}:
\begin{subequations}
\begin{align}
&\partial_i\F_{jk}+\partial_j\F_{ki}+\partial_k\F_{ij}=0\,,
\label{max1}\\
&2\,\partial_j\G_{ki}-\partial_i\G_{kj}-\partial_k\G_{ij}=
\partial_i\F_{kj}-\partial_k\F_{ji}\,,
\end{align}
\end{subequations}
with  
\begin{subequations}
\begin{align}
&\F_{ij}=\partial_j\rho_i-\partial_i\rho_j
=\partial_i\A_j-\partial_j\A_i\,,
\label{set1}\\
&\G_{ij}=\partial_i\rho_j+\partial_j\rho_i\equiv
\partial_i\A_j+\partial_j\A_i
\mod(\rho_0,\boldsymbol{\rho}_1)\,.
\end{align}
\label{FG}
\end{subequations}
The PDE's \eqref{max1} with \eqref{set1}  might be the
first set of {\it Maxwell equations}.\par Then we give a
few definitions to go further.
\begin{definition} We denote:\par
\begin{enumerate}
\item $\theta_\m$ the sheaf of rings of germs of
the differential (e.g.
$C^\infty$) functions defined on $\m$, 
\item $\underline{J_1}$ the sheaf  of $\theta_\m$-modules
of germs of differential sections of the 1-jet affine space
bundle
$J_1$ of the real valued differential functions defined on
$\m$,
\item $\Sz\subset \theta_\m$ the sheaf  of rings
of germs of solutions of the  ``c" system of algebraic
equations \eqref{third} at each point $x_0$ in $\m$
(not simultaneously at all point in $\m$; see remark
below),
\item $\Su\subset\underline{J_1}$ the sheaf of
$\theta_\m$-modules of germs of differential sections of
$J_1$ defined by the  system of algebraic equations at
each point $x_0\in\m$ (not everywhere, as mentioned
above; $i,j,k=1,\cdots,n$):
\begin{multline}
\qquad\Big\{k_0(1-\e^{2\rho_0})-
\sum_{r=1}^n\rho^r\rho_r
\,\Big\}
(\partial_k\omi{i}{j}-\partial_j\omi{i}{k})+
2k_0(\omi{i}{k}\rho_j-\omi{i}{j}\rho_k)+\\
\sum_{\ell,h=1}^n\rho^h\rho^\ell(\omi{i}{j}
\partial_k\omi{h}{\ell}-\omi{i}{k}
\partial_j\omi{h}{\ell})=0\,,
\label{manic}
\end{multline}
satisfied by
$\rho_0$ and $\boldsymbol{\rho}_1$ and deduced from the
relations \eqref{IAB} and \eqref{JAB} when $\A$ and $\B$
satisfy the relations \eqref{rhoAB},
\item $\tsms$ the sheaf of  
$\theta_\m$-modules of germs at each point $x_0\in\m$ of
global 1-forms on $\m$.
\end{enumerate}
\end{definition}
\remark We do not consider in this set
of definitions, solutions  of   PDE's but solutions of
algebraic equations at $x_0$, since  a solution of a PDE's
is a particular ``coherent" subsheaf (graphs of
solutions) for which equations \eqref{third} are
satisfied everywhere in $\m$, and not only at a given
$x_0$ whatsoever. More precisely, for instance, the
tensors $\A$ and $\B$ in \eqref{deformpar} must be such
solutions subsheafs  of the PDE \eqref{ABIJeq}. Indeed
$x_0$ may vary with these tensors  being always
solutions of the corresponding set of algebraic
equations at every point $x_0$ of subsets containing the
paths of integration (the paths can be ``cut" and the
algebraic equations must be satisfied at each new
resulting initial point of integration).\par\medskip
\remark The equations \eqref{manic}, associated to a 
particular leaf in $J_1$, define what we
call the \emph{``characteristic manifold"}.  Each
solution of \eqref{third} satisfies this set of
equations everywhere on $\m$.\par\medskip
Then, in considering the local diffeomorphisms
\(\wedge^k\,T^\ast\!\m\otimes_\Rset\!J_r
\simeq(\{x_0\}\otimes_\Rset
J_r)\times(\wedge^k\,T^\ast_{\!x_0}\!\m
\otimes_\Rset\!J_r)\) with $0\leq k\leq n$ and $r\geq0$,
we set the definitions:
\begin{definition}We define the operators:\par
\begin{enumerate}
\item
$j_1:(x_0,\rho_0)\in\Sz\longrightarrow
(x_0,\rho_0,\rho_1=\partial_1\rho_0,\cdots,\rho_n=
\partial_n\rho_0)\in\Su$,
\item 
$D_{1,c}:\boldsymbol{\rho}^1_0\equiv(x_0,\rho_0,
\boldsymbol{\rho}_1)\in\Su
\longrightarrow(\boldsymbol{\rho}^1_0,\sg_0,
\sg_1,\cdots,\sg_n)
\in\tsms\otimes_{\theta_\m}
\underline{J_1}$,
with $\A$, $\B$ and $\boldsymbol{\rho}^1_0$ satisfying
relations (\ref{rhoAB}),
\item 
\(D_{2,c}:(\boldsymbol{\rho}^1_0,\sg_0,\sg_1,\cdots,\sg_n)
\in\tsms\otimes_{\theta_\m}
\underline{J_1}\longrightarrow
(\boldsymbol{\rho}^1_0,\zeta_0,\zeta_1,\cdots,\zeta_n)\in\)\linebreak
\(\wedge^2\tsms\otimes_{\theta_\m}~\underline{J_1}\), with
\begin{align*}
&\zeta_0=\sum_{i,j=1}^n\I_{i,j}\,dx_0^i\wedge\,dx_0^j\,,\\
&\zeta_k=\sum_{i,j=1}^n\J_{j,i,k}\,dx_0^i\wedge\,dx_0^j\,,
\end{align*}
and the functions $\boldsymbol{\rho}^1_0$ and the tensors
$\I$,
$\J$,
$\A$ and $\B$  satisfying the relations (\ref{ABIJeq}).
\end{enumerate}
\end{definition}
Then from all the previous results we easily deduce:
\begin{theorem} 
The differential sequence
\begin{equation*}
\begin{CD}
0@>>>\Sz@>j_1>>\Su
@>D_{1,c}>>\tsms\otimes_{\theta_\m}\!
\underline{J_1}@>D_{2,c}>>
\wedge^2\tsms\otimes_{\theta_\m}\!\underline{J_1}\,,
\end{CD}
\end{equation*}
is  exact, and
the differential operators $D_{1,c}$ and $D_{2,c}$ are
$\Rset$-linear.
\end{theorem}
\remark We think this sequence is perhaps closed to a Spencer
non-linear sequence for the deformations of the Lie structures of the ``c" system. 
Indeed, since the system $P_c$ is integrable, it is always, at least
locally, diffeomorphic to an integrable subset of the set of Cartan 1-forms in $T^\ast\!\m
\otimes_{\Rset}\!{J_1}$
associated to a particular finite Lie algebra $g_c$ (of dimension greater or equal to
$n+1$), with corresponding Lie group
$G_c$ acting (freely or not) on the left on each characteristic manifold. Then,
$\Su$ and
$\Sz$ would be diffeomorphic  respectively (at least locally) to a sheaf of Lie groups
$G_c$, from which the Cartan forms would be defined with the first
non-linear Spencer operator, and the isotropy subgroups sheaf. Also to
$D_{2,c}$ would correspond Bianchi type equations, according to the
Nijenhuis-Fr\"olicher bracket in $T^\ast\!\m\otimes_\Rset{J_1}$, and defining the second
non-linear Spencer operator.\par The tensors
$\A$,
$\B$, $\I$ and $\J$ above are given {\it ``at the
target"\/} and not {\it ``at the source"\/}. Usually, at
the k-jets affine bundles level, the source is the point
$x_0$ and the target the point $a_0$. But at the sheafs
level, since we consider maps and not points, the source
is the
$id$ application at $x_0$, {\it e.g.} $(x_0,id)$ and the
target is the application $\hat{f}$ at $x_0$, {\it e.g.}
$(x_0,\hat{f})$, with $\hat{a}_0=\hat{f}(x_0)$. On the
other side, from the formal series point of view, the
source is $x_0$ together with the set of Taylor
coefficients of
$id$ at $x_0$ and the target is $x_0$ together with
$\{\hat{a}\}$. From the composition law of two
applications $\hat{f}\circ\hat{g}$, we can define a left
action of $\{\hat{a}\}$ on the corresponding set of
Taylor coefficients of the application $\hat{g}$. Since
the conformal Lie equations are involutive, then we
obtain a left action of  a finite set of Taylor
coefficients of order less than 3. Hence the set of
values taken by this finite set define a Lie group
we denote $\widehat{G}_{2,x_0}$ at $x_0$. Moreover its left
action links any source point $(x_0,\{\hat{a}'\})$  to
any other $(x_0,\{\hat{a}''\})$. In fact, it 
indicates  the property of transitivity of the
conformal Lie groupoid defined by the system
\eqref{confeq}.  Hence, its action on the source defines
a left action of a transitive finite Lie group on the
leafs of $J_1$ (not
$\m$), as well as a similar transitive left action on
$\widehat{G}_{2,x_0}$ itself. Then this group is diffeomorphic to a subgroup of the
previously defined
$G_c$ group. It follows from the transitivity, inducing globality, and in case of a compact
manifold
$\m$, the integrals in
\eqref{deformpar} would define perhaps a deformation class in the first
Spencer cohomology space of deformations of global sections from $\m$ to the sheaf of
$\widehat{G}_{2}$ Lie groups.\par   
In the ``diagonal method
approach" used by A. Kumpera
\& D. Spencer \cite{kumperaspencer}, formulas 
are given at the source  by using the {\it Buttin
formula}. It is out of our purpose and not necessary to
give an equivalent formula in the present formalism. The results at the source will be
equivalent to those at the target because of the transitivity.\par
We can give, as an example, what is this action 
of $\widehat{G}_2$ on the $c$'s. From
\eqref{albet}, one has the relations, with
$\sum_{j=1}^n\hat{b}^i_j\,\hat{a}^j_k=
\sum_{j=1}^n\hat{a}^i_j\,\hat{b}^j_k=\delta^i_k\,$:
\begin{align*}
&{c'}_0=c_0-\frac{1}{n}\ln{|}\det
(\hat{a}^i_j)\,{|}\,,\\
&\sum_{i=1}^{n}\hat{a}^i_j\,{c'}_i=c_j-\sum_{k,l=1}^{n}
\frac{1}{n}\,\hat{b}^{k}_l\,\hat{a}^l_{kj}\,.
\end{align*}
\remark In view of physical
interpretations, we compute the Euler-Lagrange equations
of a Lagrangian density
\begin{equation}
\Lag(x_0,\rho_0,\boldsymbol{\rho}_1,\A,\F,\G)\,d^n\!x_0\,,
\label{LAB}
\end{equation}
with $\A$, $\F$ and $\G$ satisfying the relations
\eqref{rhoA} and \eqref{FG}. We obtain what we call {\it
the second set of differential equations}:
\begin{subequations}
\begin{align}
&\sum_{k=1}^n\partial_k\left(
\frac{\partial\Lag}{\partial\A_k}\right)=
\left(\frac{\partial\Lag}{\partial\rho_0}\right)\,,
\label{cur}\\
&\sum_{k=1}^n\left\{\partial_k\left(
\frac{\partial\Lag}{\partial\F_{ik}}\right)+
\partial_k\left(
\frac{\partial\Lag}{\partial\G_{ik}}\right)\right\}=
\frac{1}{2}\left\{
\left(\frac{\partial\Lag}{\partial\rho_i}\right)-
\left(\frac{\partial\Lag}{\partial\A_i}\right)
\right\}\,.
\label{max2}
\end{align}
\end{subequations}
If we consider $\A$ as being the electromagnetic potential
vector, $\F$ the Faraday tensor and $\Lag$ only depending
on $\A$ and $\F$, then \eqref{max1} and \eqref{max2} are
the {\it second set of Maxwell equations} if we denote by
\begin{equation*}
\J^k_e=\left(\frac{\partial\Lag}{\partial\A_k}\right)\,,
\end{equation*}
the electric current components and
\begin{equation*}
\Pol^{ki}=-2\,\left(
\frac{\partial\Lag}{\partial\F_{ki}}\right)\,,
\end{equation*}
the polarization tensor components. Moreover the
differential equations \eqref{cur} become the free
divergence property of the electric current $\J_e$ if
$\Lag$ is independent of $\rho_0$. It has to be noticed
that no magnetic currents seem to occur.
\subsection{The ``full" system} This system is defined
by the set of PDE's \eqref{confeq}. 
For this system of Lie equations, we recall well-known
results but in the present context. Applying the same
reasoning than in the previous subsection, firstly we
obtain the following results up to order two: 
\begin{subequations}
\begin{align}
&\sum_{r,s=1}^n\omega_{rs}(\hat{a}_0)
\,\,\hat{a}^{r}_{i}\,\hat{a}^{s}_{j}=
e^{2c_0}\omega_{ij}(x_0)\,,
\label{a1h}\\ 
&\hat{a}^k_{ij}+
\sum_{r,s=1}^n{}\gamma^{k}_{rs}(
\hat{a}_0)\,\,\hat{a}^{r}_{i}\,\hat{a}^{s}_{j}=
\sum_{q=1}^n{}\hat{a}_q^k\left(
\gamma^{q}_{ij}(x_0)+c_i\delta^q_j+
c_j\delta^q_i-\omega_{ij}(x_0)c^q\right)\,,
\label{a2h}
\end{align}
\label{a12h}
\end{subequations}
which clearly shows that $J_1$ is
diffeomorphic to an embedded submanifold of the 2-jets
affine bundle $J_2(\m)$ of the $C^\infty(\m,\m)$
differentiable applications on
$\m$. Secondly we get relations for the
coefficients of order 3 that we only write as
($\hat{a}_1\equiv(\hat{a}^i_j)$;
$\hat{a}_2\equiv(\hat{a}^i_{jk})$,\dots,
$\hat{a}_k\equiv(\hat{a}^i_{j_1\cdots j_k})$;
$\bfah^k_0\equiv(\hat{a}_0,\cdots,\hat{a}_k)$):
\begin{equation}
\hat{a}^i_{jkh}\equiv
\hat{A}^i_{jkh}(x_0,\bfah^2_0)\,,
\label{a3h}
\end{equation}
pointing out in this expression the independence from
the ``$c$" coefficients. We denote by
$\Oc^i_J$ the Pfaff 1-forms at $x_0$ and $\{\hat{a}\}$
(or at the target $(x_0,\{\hat{a}\})$):
\begin{equation}
\Oc^i_J\equiv
d\hat{a}^i_J-\sum_{k=1}^n\hat{a}^i_{J+1_k}dx^k_0\,,
\label{Oma}
\end{equation}
and  setting the $\hat{a}$'s as values of
functions $\hat{\tau}$ depending on $x_0$ (in some way we
make a pull-back on $\m$), we define the tensors $\kc$ 
and the differential operator $\widehat{\text{\eu
D}}_1$ by:
\begin{equation}
\Oc^i_J\equiv\widehat{\text{\eu
D}}_1\hat{\tau}^i_J\equiv\sum_{k=1}^n\Big(\partial_k\hat{\tau}^i_J-
\hat{\tau}^i_{J+1_k}\Big)\,dx^k_0\equiv
\sum_{k=1}^n\kc^i_{k,J}\,
dx^k_0\,.
\label{omtau}
\end{equation}
At this step, we can have a look on the
corresponding 1-forms ${}^s\Oc$ at the source 
$(x_0,id)$. Using the composition law
$\hat{f}''\equiv\hat{f}\circ\hat{f}'$ of two
diffeomorphisms $\hat{f}$ and $\hat{f}'$, we deduce from
the relations
\begin{align*}
&\hat{f}''{}^i_j=\sum_{k=1}^n\hat{f}^i_k\circ\hat{f}'
\;\hat{f}'{}^k_j\,,\\
&\hat{f}''{}^i_{jk}=
\sum_{r,s=1}^n\hat{f}^i_{rs}\circ\hat{f}'
\;\hat{f}'{}^r_j\hat{f}'{}^s_k+
\sum_{u=1}^n\hat{f}^i_u\circ\hat{f}'\hat{f}'{}^u_{jk}\,,\\
&\cdots=\cdots\,,
\end{align*}
and their Taylor coefficients versions, the relations
for the 1-forms ${}^s\Oc$:
\begin{align*}
&{}^s\Oc^k=\sum_{i=1}^n\;\hat{b}^k_i\;\Oc^i\,,\\
&{}^s\Oc^k_j=\sum_{i=1}^n\;\hat{b}^k_i\left\{\Oc^i_j-
\sum_{s,r=1}^n\;\hat{a}^i_{js}\hat{b}^s_r\;\Oc^r
\right\}\,,\\
&\cdots=\cdots\,.
\end{align*}
Expressing the latter with the 1-forms $dx_0^i$ and
$d\hat{a}^k_J$, we recover similar expressions than
those given by J.-F. Pommaret in \cite{pomm94} p.214 for
the ``$\chi$ tensors" but, from our opinion, in a clearer
context:
\begin{align*}
&{}^s\Oc^i=\sum_{j=1}^n\hat{b}^i_j\;d\hat{a}^j-dx_0^i\,,\\
&{}^s\Oc^i_j=\sum_{k=1}^n\hat{b}^i_k
\left\{d\hat{a}^k_j-\sum_{r,s=1}^n\hat{a}^k_{jr}
\hat{b}^r_sd\hat{a}^s
\right\}\,,\\
&\cdots=\cdots\,.
\end{align*}
These 1-forms are conformal invariant Cartan 1-forms. We think it also gives an other
presentation of the {\it``Buttin formula"\/}. Then from
the relations:
\begin{align*}
&\e^{2c_0}\oms{r}{s}(\hat{a}_0)=
\sum_{i,j=1}^n\oms{i}{j}(x_0)\hat{a}^r_i
\hat{a}^s_j\,,\\
&\sum_{i=1}^n\gam{i}{i}{k}=
\frac{1}{2}\sum_{i,j=1}^n\oms{i}{j}\,\partial_k\omi{i}{j}\,,
\end{align*}
we deduce, for example, the $\Oc^i_j$ 1-forms
satisfy the relations at
the target
$(x_0,\bfah^1_0)$:
\begin{equation}
\widehat{H}_0(x_0,\bfah^1_0,\Oc^k_L;|L|\leq1)
\equiv\sum_{i,j=1}^n\hat{b}^j_i\,\Oc^i_j+
\sum_{j,k=1}^n\gamma^j_{jk}(\hat{a}_0)\Oc^k
=n\sg_0\,.\label{nsg0}
\end{equation}
Similar computations show that the 1-forms $\sg_i$
can be expressed as quite long relations, linear in the
$\Oc^j_J$ ($|J|\leq2$) and with polynomials of the
$\hat{a}_K$ ($|K|\leq2$) and derivatives of the metric
and the Riemann-Christoffel symbols taken either at $x_0$ or
$\hat{a}_0$, as coefficients. Then, we set:
\begin{equation}
\sg_i\equiv\widehat{H}_i(x_0,\bfah^2_0,\Oc^j_I;
|I|\leq2)\,.
\label{nsgi}
\end{equation}
From \eqref{a3h} the 1-forms 
$\Oc^i_{jkh}$  are also sums of 1-forms
$\Oc^r_K$ ($|K|\leq2$) with the same kind of coefficients
and not depending on the $\sg$'s, and we write:
\begin{equation}
\Oc^i_{jkh}\equiv\widehat{K}^i_{jkh}(x_0,
\bfah^2_0,\Oc^r_K;|K|\leq2)\,,
\label{K3}
\end{equation}
where $\widehat{K}^i_{jkh}$ is linear in the 1-forms
$\Oc^r_K$.\par\medskip
Denoting by $\text{\eu P}_2\subset J_2(\m)$ the set of
elements satisfying relations \eqref{a12h} whatever are
the $c$'s.  Then the Pfaff system we denote
$\widehat{P}_2$ over $\text{\eu P}_2$ and generated by
the 1-forms $\Oc^j_K\in{T^\ast\!\m}
\otimes_{\Rset}\!{J_2(\m)}$
in \eqref{Oma} with $|K|\leq2$, is locally integrable on
every neighborhood
$U_{(x_0,\bfah_0^2)}{\subset}J_2(\m)$ since at the
target $(x_0,\bfah^2_0)$ we have ($|J|\leq2$):
\begin{equation}
\Tc^i_J\equiv\widehat{\text{\eu D}}_2\Oc^i_J\equiv
d\Oc^i_J-\sum_{k=1}^n dx^k_0\wedge\Oc^i_{J+1_k}=0\,,
\label{dom}
\end{equation}
together with \eqref{K3}.\par\medskip
From now we consider the ``Poincar\'e system" with
corresponding notation without the ``hats". We denote by
$\Omega^i_J$ the Pfaff 1-forms corresponding to this
system, {\it e.g.} the system defined by the PDE's
\eqref{f1} and \eqref{f2} with a vanishing function
$\alpha$. The corresponding 1-forms ``$\sg$"  are also
vanishing everywhere on $\m$ and the $\Omega^i_J$
satisfy all the previous relations but with the $\sg$'s
cancelled out. Then it is easy to
see the 
$\Omega^i_J$ 1-forms ($|J|\geq2$) are generated
by the set of 1-forms $\Omega^j_K$ ($|K|\leq1$)
and in particular we have 
\begin{equation}
\Omega^k_{ij}=-\left\{\sum_{r,s,h=1}^n(\partial_h
\gamma^h_{ij})({a}_0)\Omega^r\,{a}^r_i\,{a}^s_j
+\sum_{r,s=1}^n\gamma^k_{rs}({a}_0)[\,{a}^r_i\,\Omega^s_j+
{a}^s_j\,\Omega^r_i\,]\right\}\,,
\label{K2}
\end{equation}
with $\bfah_0^1\equiv \bfa_0^1\in\text{\eu P}_1\subset
J_1(\m)$, and $\text{\eu P}_1$ being the set of elements
satisfying relations \eqref{a1h} with $c_0=0$. Similarly
the Pfaff system we denote ${P}_1$ over $\text{\eu P}_1$
and generated by the 1-forms
$\Omega^j_K$ in \eqref{Oma} with $|K|\leq1$, is
locally integrable on every neighborhood
$U_{(x_0,\bfa_0^1)}\subset\text{\eu P}_1$ since at the
$(x_0,\bfa_0^1)$ point we have the relations
\eqref{dom} with $|J|\leq1$ together with the relations
\eqref{K2}.\par
Then we have at each  $(x_0,\bfah_0^2)\in\text{\eu P}_2$
the locally exact splitted sequence
\begin{equation}
\begin{CD}
0@>>>P_1@>b_1>>\widehat{P}_2@>e_1>>P_c@>>>0\,,\label{pseq}
\end{CD}
\end{equation}
where we consider $J_1$ embedded in $J_2(\m)$ as well
as  $\text{\eu P}_1$ from relations \eqref{a12h}. In
this sequence a back connection
$b_1$  and a connection
$c_1:P_c\longrightarrow\widehat{P}_2$ are such that
($|J|\leq2$):
\begin{equation}
\Oc^i_J=\Omega^i_J+\chi^i_J(\bfa_0^2)\,\sg_0+
\sum_{k=1}^n\chi^{i,k}_J(\bfa_0^2)\,\sg_k\,,
\label{splitom}
\end{equation}
with $\Omega^i_{jk}$ satisfying \eqref{K2} for any given 
$\Omega^h_L$ with $|L|\leq1$, and the tensors $\chi$ defined on $\text{\eu
P}_2$. They define together a back connection, and  the tensors
$\chi$ define a
connection if they satisfy the relations:
\begin{subequations}
\begin{align}
&\widehat{H}_0(x_0,\bfah^1_0,\chi^k_L;|L|\leq1)=n\,,\\
&\widehat{H}_0(x_0,\bfah^1_0,\chi^{k,i}_L;|L|\leq1)=0\,,\\
&\widehat{H}_i(x_0,\bfah^2_0,\chi^k_L;|L|\leq2)=0\,,\\
&\widehat{H}_i(x_0,\bfah^2_0,\chi^{k,h}_L;|L|\leq2)=
n\delta^h_i\,,
\end{align}
\label{lesH}
\end{subequations}
in order to keep the relations \eqref{nsg0} and 
\eqref{nsgi}, {\it e.g.} $e_1\circ c_1=id$.\par\medskip
\remark The corresponding {\it ``characteristic
manifolds"\/} for the conformal and Poincar\'e Pfaff
systems are defined by tremendous sets of algebraic
equations.  We  give rather a way of computations to get
them if really necessary (!) but especially to see their
algebraic features in the k-jets affine bundles on $\m$.
First of all, as it can easily be verified, these
algebraic equations come from the ($k+1$)th-order. For
instance, in the conformal case, we begin with the
equations
\eqref{a3h} of order three since the conformal Pfaff
system is defined on
$J_2(\m)$. Then we easily deduce a first set of relations
of the kind:
\begin{equation*}
\Oc^i_{jkh}=\sum^n_{r=1}\sum_{|L|\leq2}
\left(\frac{\partial\widehat{A}^i_{jkh}(x_0,\boldsymbol{\hat{\tau}}^2_0)}
{\partial\hat{\tau}^r_L}\right)\,\Oc^r_L\,,
\end{equation*}
where we substitute the values $\hat{a}$ by the
functions $\hat{\tau}$ satisfying the
relations \eqref{a3h} again. From
\eqref{omtau} with $|J|=2$, we finally obtain the
algebraic equations of the characteristic manifold  as
expressions such as
\begin{equation*}
\sum^n_{j,k=1}\left\{
\left(\frac{\partial\widehat{A}^i_{rsk}}
{\partial x^j_0}\right)+
\sum^n_{u=1}\sum_{|L|\leq2}
\left(\frac{\partial\widehat{A}^i_{rsk}}
{\partial\hat{\tau}^u_L}\right)\hat{\tau}^u_{L+1_j}
\right\}\,dx^j_0\wedge dx^k_0=0\,,
\end{equation*}
into which the $\hat{\tau}^u_K$ functions of order three
($|K|=3$) are expressed by  functions $\hat{\tau}$ of
less order with relations \eqref{a3h}. It underlines the
skew-symmetry property of these
characteristic manifolds which appear to
be symmetric spaces. 
\section{The differential sequences}
Let us denote by $\Pi_k(\m)$ the subbundle of $J_k(\m)$
of the local diffeomorphisms of $\m$, and
$j_k:(x^j_0,\hat{\tau}^i)\in
\underline{J_0(\m)}\longrightarrow(x_0^j,
\hat{\tau}^i,\partial_j\hat{\tau}^i,\cdots,\partial_{j_1\cdots
j_k}\hat{\tau}^i)
\in\underline{J_k(\m)}$, where we underline the
$k$-jets affine bundles to indicate their corresponding
sheafs as usual. As for the ``$c$" system we give the
following set of definitions.
\begin{definition}
We denote:
\begin{enumerate}
\item $\Gamma_0(\m)\subset\underline{\Pi_0(\m)}$ the
sheaf of
$\theta_\m$-modules of germs of solutions of the
Poincar\'e system of algebraic equations \eqref{f1} and
\eqref{f2} with $\alpha=0$ at each point $x_0$ in $\m$,
\item $\Gamma_1(\m)\subset\underline{\Pi_1(\m)}$ the
sheaf of $\theta_\m$-modules of germs of differential
sections of $\,\Pi_1(\m)$ satisfying  the
\emph{``characteristic manifold"\/} algebraic equations
for the\linebreak Poincar\'e system at each $x_0\in\m$,
and which inherits of the \emph{``Lie groupoid
structure"\/} from
$\,\Pi_1(\m)$.
\end{enumerate}
\end{definition}
\begin{definition}
We denote:
\begin{enumerate}
\item
$\widehat{\Gamma}_0(\m)\subset\underline{\Pi_0(\m)}$ the
sheaf of $\theta_\m$-modules of germs of solutions of the
conformal system of algebraic equations \eqref{f1} and
\eqref{f2}, whatever is $\alpha$, and extended by the
\emph{``third order system"\/} at each point $x_0$ in
$\m$,
\item
$\widehat{\Gamma}_2(\m)\subset\underline{\Pi_2(\m)}$ the
sheaf of
$\theta_\m$-modules of germs of differential sections of
$\,\Pi_2(\m)$ satisfying  the \emph{``characteristic
manifold"\/} algebraic equations for the conformal system
at each
$x_0\in\m$, and which inherits of the \emph{``Lie
groupoid structure"\/} from $\,\Pi_2(\m)$.
\end{enumerate}
\end{definition}
With the local diffeomorphisms 
\(\wedge^k\,T^\ast\!\m
\otimes_\Rset\!J_r(\m)
\simeq(\{x_0\}\otimes_\Rset
\!J_r(\m))\times\linebreak(\wedge^k\,T^\ast_{\!x_0}\!\m
\otimes_\Rset\!J_r(\m))\), and from the last section we
deduce the theorems:
\begin{theorem}
The differential sequences below, where in the
differential operators
$\widehat{\text{\eu D}}_1$, $\widehat{\text{\eu D}}_2$, 
${\text{\eu D}}_1$ and  ${\text{\eu D}}_2$ are
$\Rset$-linear,
\begin{equation*}
\begin{CD}
0@>>>\Gamma_0@>{j_1}>>
\Gamma_1@>{\text{\eu D}}_1>>\underline{T^\ast\!\m}
\otimes_{\theta_\m}\!\underline{J_1(\m)}@>
{\text{\eu D}}_2>>
\wedge^2\underline{T^\ast\!\m}
\otimes_{\theta_\m}\!\underline{J_1(\m)}\,,
\end{CD}
\end{equation*}
\begin{equation*}
\begin{CD}
0@>>>\widehat{\Gamma}_0@>{j_2}>>
\widehat{\Gamma}_2@>\widehat{\text{\eu
D}}_1>>\underline{T^\ast\!\m}
\otimes_{\theta_\m}\!\underline{J_2(\m)}@>
\widehat{\text{\eu D}}_2>>
\wedge^2\underline{T^\ast\!\m}
\otimes_{\theta_\m}\!\underline{J_2(\m)}
\end{CD}
\end{equation*}
are exact.
\end{theorem}
\remark In each of these theorems, the sheafs
$\underline{\Pi_i(\m)}$ ($i=1,2$) in the source
sheafs of the differential operators ${\text{\eu D}}_2$
and $\widehat{\text{\eu D}}_2$, must be taken into
account in a similar way as for
$\underline{J_1}$ in the corresponding sequence for the
``$c$" system.\par\medskip
From all the previous theorems and the splitting
\eqref{pseq} and the one defined by \eqref{a12h} we also deduce:
\begin{theorem}
The following diagram
\vfill\eject
\[\begin{CD}
@.1@.1@.0@.\\
@.@VVV@VVV@VVV@.\\
0@>>>\Gamma_0@>{j_1}>>
\Gamma_1@>{\text{\eu D}}_1>>\underline{T^\ast\!\m}
\otimes_{\theta_\m}\underline{J_1(\m)}@>
{\text{\eu D}}_2>>
\wedge^2\underline{T^\ast\!\m}
\otimes_{\theta_\m}\!\underline{J_1(\m)}\\
@.@VVV@VV\underline{b_0}V@VV
\underline{b_1}V@.\\
0@>>>\widehat{\Gamma}_0@>{j_2}>>
\widehat{\Gamma}_2@>\widehat{\text{\eu
D}}_1>>\underline{T^\ast\!\m}
\otimes_{\theta_\m}\!\underline{J_2(\m)}@>
\widehat{\text{\eu D}}_2>>
\wedge^2\underline{T^\ast\!\m}
\otimes_{\theta_\m}\!\underline{J_2(\m)}\\
@.@.@VV\underline{e_0}V@VV\underline{e_1}V
@.\\
0@>>>\Sz@>j_1>>\Su
@>D_{1,c}>>\tsms\otimes_{\theta_\m}\!
\underline{J_1}@>D_{2,c}>>
\wedge^2\tsms\otimes_{\theta_\m}\!\underline{J_1}\\
@.@.@VVV@VVV@.\\
@.@.0@.0@.
\end{CD}\]
is a commutative diagram of exact sequences of sheafs of
modules with $\Rset$-linear differential operators.
\end{theorem}
\remark We do not have  any relation of the kind
\(\underline{e_2}\circ\widehat{\text{\eu
D}}_2=D_{2,c}\circ\underline{e_1}\), where
$\underline{e}_2$ would be a map from \(\underline{T^\ast\!\m}
\otimes_{\theta_\m}\!\underline{J_2(\m)}\) to 
\(\underline{T^\ast\!\m}
\otimes_{\theta_\m}\!\underline{J_1}\). Hence
the commutativity in the latter diagram can't be
``extended" on the right to give a right vertical
sequence between the sheafs of 2-forms. Indeed,
in considering the splitting \eqref{splitom} in
the expression \(d\Oc^i_J-\sum_{k=1}^n
dx^k_0\wedge\Oc^i_{J+1_k}\), and setting ($r=1,\cdots,n$)
\begin{align*}
&\mu_0=d\sg_0-\sum_{k=1}^ndx_0^k\wedge\sg_k\,,\\
&\mu_r=d\sg_r-\sum_{k=1}^ndx_0^k\wedge\sg_{kr}\,,
\end{align*}
we obtain:
\begin{multline*}
d\Oc^i_J-\sum_{k=1}^n dx^k_0\wedge\Oc^i_{J+1_k}=
d\Omega^i_J-\sum_{k=1}^n
dx^k_0\wedge\Omega^i_{J+1_k}+
\chi^i_j\mu_0+\sum_{k=1}^n\chi_J^{i,k}\mu_k+\\
\hspace{4cm}[d\chi^i_J-\sum_{k=1}^n\chi^i_{J+1_k}dx_0^k]\wedge\sg_0+
\sum_{k,r=1}^n\chi^{i,k}_J
dx_0^r\wedge\sg_{kr}+\\
\sum_{k=1}^n[d\chi^{i,k}_J-
\sum_{r=1}^n(\chi^{i,k}_{J+1_r}-
\chi^i_{J}\delta_r^k)dx_0^r]
\wedge\sg_k\,.
\end{multline*}
Then, if the relations \eqref{K3} are satisfied
(which are also  satisfied by the $\Omega^i_J$ with
$|J|\leq3$), together with the relations \eqref{K2} and
\eqref{sg2}, we deduce
\begin{multline*}
\widehat{\text{\eu D}}_2\Oc^i_J=\text{\eu
D}_2\Omega^i_J+
[d\chi^i_J-\sum_{k=1}^n\chi^i_{J+1_k}dx_0^k]\wedge\sg_0+
\sum_{k=1}^n[d\chi^{i,k}_J-
\sum_{r=1}^n(\chi^{i,k}_{J+1_r}-
\chi^i_{J}\delta_r^k)dx_0^r]
\wedge\sg_k+\\
\sum_{k,r=1}^n\chi^{i,k}_J
dx_0^r\wedge\vartheta_{kr}(\bfc_0^1,\sg_J)+
\chi^i_j\,D_{2,c}\sg_0+
\sum_{k=1}^n\chi_J^{i,k}\,D_{2,c}\sg_k\,.
\end{multline*}
In order to make the diagram commutative then we must
set:
\begin{multline*}
[d\chi^i_J-\sum_{k=1}^n\chi^i_{J+1_k}dx_0^k]\wedge\sg_0+
\sum_{k,r=1}^n\chi^{i,k}_J
dx_0^r\wedge\vartheta_{kr}(\bfc_0^1,\sg_J)\\
+\sum_{k=1}^n[d\chi^{i,k}_J-
\sum_{r=1}^n(\chi^{i,k}_{J+1_r}-
\chi^i_{J}\delta_r^k)dx_0^r]
\wedge\sg_k=0\,.
\end{multline*}
It follows from the latter equations that
\begin{align*}
&d\chi_J^i\wedge dx_0^1\wedge\cdots\wedge
dx_0^n\wedge\sg_0\wedge\sg_1\wedge\cdots\wedge\sg_n=0\,,\\
&d\chi_J^{i,k}\wedge dx_0^1\wedge\cdots\wedge
dx_0^n\wedge\sg_0\wedge\sg_1\wedge\cdots\wedge\sg_n=0\,,
\end{align*}
which shows clearly that \(d\chi_J^{i,k}\) and
\(d\chi_J^{i}\)  are defined on $T^\ast\!J_1$, and
consequently the $\chi$'s  depend on
$\bfc_0^1$. But from the relations \eqref{lesH} the
$\chi$'s also depend on $\bfah_0^2$, which involves that
a sequence of sheafs between the sheafs of 2-forms would
exist only if a section from $J_1$ to $\text{\eu P}_2$
is given. Anyway in this case the bottom and the middle
sequences of the last theorem would be merely isomorphic.
\section{The unfolded space-time and the gravitation}
Again, in view of physical interpretations, we put a
spotlight on the tensor $\B$. In fact, we
consider the relations \eqref{splitom} with $|J|=0$ and
the $\Oc^i$ as fields of {\em``tetrads"\/} or
{\em``soldering 1-forms"\/}. Then we get a new metric
$\nu$ of what we call the {\it``unfolded space-time"\/}
defined at $x_0$ by:
\begin{align*}
&\nu=\sum_{i,j=1}^n\omega_{ij}(x_0)\,
\Oc^i(x_0)\otimes\Oc^j(x_0)\,,\\
&\Oc^i=\sum_{k=1}^n\kc^i_{k}(x_0)\,
dx^k_0\,,\\
&\kc^i_j=\kappa^i_j(x_0)+\chi^i(x_0,\bftauh_0^2)\,\A_j(x_0)+
\sum_{k=1}^n\chi^{i,k}(x_0,\bftauh_0^2)\,\B_{j,k}(x_0)\,.
\end{align*}
We consider the particular case for which the metric
$\omega\equiv\text{diag}[+,-,\cdots,-]$, the
$\chi$'s are only depending on $x_0$  and
$\kappa^i_j=\delta^i_j$, {\it e.g.} the deformation of
$\omega$ is only due to the tensors $\A$ and
$\B$. Thus, one has the general relation between
$\nu$ and $\omega$: $\nu=\omega\,+$ linear and quadratic
terms in $\A$ and $\B$. Then from this metric
$\nu$, one can deduce the Riemann and Weyl curvature
tensors of the {\it``unfolded space-time"\/}. One has a
non-metrical theory for the gravitation in the gauge
space-time, since clearly $\nu$  doesn't appear  as a
gravitational potential.\par The space-time terminology
we use is quite	natural in the sense that one has
simultaneously two types of space-time. The first one,
which we call the {\em ``underlying"\/} or {\em
``substrat" space-time\/}, is endowed	 with the metric
$\omega$ and is of constant scalar curvature $k_0$. It is
the {\it``compars" space-time\/}, {\it e.g.} the
space-time of physical or material rulers and watches, or
merely of {\it ``material bodies"\/} ({\it e.g.}
electrically charged or magnetized, massive and featured
by the weights $(m,s)$ of the finite irreducible
representations of the Poincar\'e Lie group). The other
one, called the {\it ``dispars"} or {\it ``unfolded 
space-time"\/}, endowed with  the metric
$\nu$, is defined for any scalar curvature and by the
gauge potentials $\A$ and $\B$. It can be
considered as the underlying space-time, deformed by the
gauge potentials and the Weyl curvature does not
necessarily vanish. Moreover, from a continuum mechanics
of deformable bodies point of view, the metric
$\nu$ can be interpreted as the tensor of deformation of
the underlying space-time~\cite{kata}.\par 
We are faced to a question: could this kind of deformation
be interpreted as an inflation process in
cosmology in which each occurrence (a tick-tock) of a 
creation or annihilation of a potential of local
interaction would lead to a {\it ``thermodynamic
clock"\/} related to the unfolding in the sense of I.
Prigogine (not a geometric clock, this latter being
associated to the non {\it ``topologically evolving"\/} 
substrat space-time~!)  and inducing a {\it
``measurable"\/} time evolution without measurable time
origin as a consequence ? Indeed, in such situation,
clocks would be produced by the time  and conversely !
Also, would the inflation be the evolution from the
Poincar\'e Lie structure to ``a" conformal one, and going
from a physically homogeneous space-time (namely with
constant Riemann curvature and a vanishing Weyl tensor
and so {\it ``rigid"\/}) to an inhomogeneous one (with
any Weyl tensor) ? Coming from a vanishing Weyl tensor to
a non-vanishing one, is in accordance with some Big-Bang
concepts, but with a very different meaning of time and
space-time than those that would be considered in the
aforementioned context. In some way, the potentials
$\A$ and $\B$ would produce a {\it
``bifurcation"\/} of the space-time structure leading to
a different concept of bifurcation than the one used in
the case of non-linear ODE's. A counting of
the bifurcations would be an evaluation
of a particular time, {\it e.g.} an ordered set of agreements on events shared by
observers, each one associated to a Lorentz frame.\par\medskip  In view of
making easier computations for a relativistic action deduced from the metric
tensor
$\nu$, we consider this metric in the {\it ``weak fields
limit"\/}, {\it e.g} the metric $\nu$ is linear in the
tensors $\A$  and $\B$ and the quadratic terms
are neglected. Furthermore, from the relations
\eqref{ABIJeq}, we have  the
relations:
\begin{equation*}
\partial_i\A_k-\partial_k\A_i=\B_{k,i}-\B_{i,k}=
\F_{k,i}\,,
\end{equation*}
\begin{equation*}
\partial_j\B_{k,i}-\partial_k\B_{j,i}\simeq
k_0(\omega_{ik}\A_j-\omega_{ij}\A_k)\,,
\end{equation*}
since  the functions $\rho$ take also small
values in the weak fields limit. Therefore, we can write 
\begin{equation*}
\nu_{ij}=\omega_{ij}+\epsilon_{ij}\,,
\end{equation*}
where the $\epsilon_{ij}$ coefficients are small
perturbations of the metric $\omega$ and defined from
$\A$ and $\B$ by the formula:\par
\begin{equation*}
\epsilon_{ij}\simeq
\sum_{k=1}^n\chi^k\left(\omega_{kj}\A_i+
\omega_{ki}\A_j\right)+
\sum_{k,h=1}^n\chi^{k,h}\left(\omega_{kj}\B_{i,h}+
\omega_{ki}\B_{j,h}\right)\,.
\end{equation*}
Then, let $i$ be a differential map
$i:s\in[0,\ell]\subset\Rset\longrightarrow i(s)=x_0\in\m$\,. We
define the relativistic action $S_1$ by:
\begin{equation*}
S_1=\int_0^\ell\sqrt{\nu(u(s),u(s))}\;ds
\equiv\int_0^\ell \sqrt{2L_\nu}\;ds\,,
\end{equation*}
where $u(s)\equiv di(s)/ds\,$. We also take the tensors
$\chi$ as depending on $s$. The Euler-Lagrange equations
for the Lagrangian density $\sqrt{L_\nu}$ are not
independent because $\sqrt{L_\nu}$ is a homogeneous
function of degree 1 and thus satisfies an additional
homogeneous differential equation. Then, it is well-known
that the variational problem for
$S_1$ is equivalent to consider the variation of the
action $S_2$ defined by
\begin{equation*}
S_2=\int_0^\ell{\nu(u(s),u(s))}\;ds
\equiv\int_0^\ell 2L_\nu\;ds\,,
\end{equation*}
but constrained by the condition $2L_\nu=1$. In this
case,  it shows that $2L_\nu$ must be
considered, firstly, as a Lagrange multiplier, and
secondly, its explicit expression with respect to $u$
will appear only in the variational calculus. In the weak
fields limit and with
\({\parallel}u
{\parallel}^2\equiv\omega(u,u)=\sum_{i,j=1}^n\omega_{ij}u^iu^j=
(u^1)^2-(u^2)^2-\cdots-(u^n)^2\,\), we obtain:
\begin{equation}
L_\nu=\frac{1}{2}{\parallel}u{\parallel}^2+
\sum_{j,k=1}^n\omega_{kj}\,\chi^k\,u^j\;.\;
\sum_{i=1}^n\A_i\,u^i+
\sum_{j,k,h=1}^n\chi^{k,h}\,\omega_{kj}\,u^j\;.\;
\sum_{i=1}^nu^i\,\B_{i,h}\,.
\label{lnu}
\end{equation}
But also, if $\,0<{\parallel}u{\parallel}^2<\!+\infty\,$
and setting ${\parallel}u{\parallel}\equiv1/\gamma\,$, we
have:
\begin{equation}
\sqrt{2L_\nu}\,\simeq\gamma^{-1}+
\sum_{j,k=1}^n\omega_{kj}\,\chi^k\,(\gamma u^j)\,.
\sum_{i=1}^n\A_i\,u^i+
\sum_{j,k,h=1}^n\chi^{k,h}\,\omega_{kj}\,(\gamma
u^j)\,.
\sum_{i=1}^n u^i\,\B_{i,h}\,.
\label{lnubis}
\end{equation}
From the latter relation, we can deduce a few
physical consequences among a lot of other particular
ones. On one hand, if we assume that
\begin{subequations}
\begin{align}
&\sum_{j,k=1}^n\omega_{kj}\chi^{k,h}U^j\equiv\xi^h
=cst\,,
\label{thomgen}\\
&\sum_{k,j=1}^n\omega_{kj}\chi^kU^j\equiv\xi_0=cst\,,
\label{thomas}
\end{align}
\label{prec}
\end{subequations}
where $U\equiv\gamma u$, then we recover in
\eqref{lnubis}, up to some suitable constants, the
Lagrangian density for a particle, with the {\it
$\,n$-velocity vector $U$\/}
(\({\parallel}U{\parallel}^2=1\)), embedded in an
external electromagnetic field. But also from the relation
\eqref{thomas} we find a {\it``Thomas precession"\/} if 
the tensor $(\chi^k)$ is ascribed to a
{\it``polarization
$n$-vector"\/} \cite[p.270]{bacry} {\it``dressing"\/} the
particle (ex.: the spin of an electron).\par  On the
other hand, we have also a physical interpretation in the
framework of the  magnetoelectric interaction phenomena
in crystals as described in chapter 2. Assuming  the
symmetric part of
$\B$ is vanishing only, $\xi^h=cst\neq0$ and the relation
\eqref{thomas} satisfied, then
we recover in \eqref{lnubis} the Lagrangian density
${}^2\!\mathcal{L}$ of E. Asher \cite{asher} for
magnetoelectric phenomena. Moreover the relation
\eqref{thomgen} with $\xi^h=cst\neq0$ is precisely the
condition for to get a {\it``generalized Thomas
precession"\/} in magnetoelectric crystals. This
generalized precession could give a possible origin for
the creation of anyons in high-$T_c$ superconductors
\cite{rub93,rub94} and might be an alternative to
Chern-Simon theory. In this situation the tensor
$(\chi^{k,h})$ is a polarization tensor of a crystal
and the particle is {\it``dressed"\/} with this kind of
polarization (not polarized by\dots). The condition
$\xi^h=cst$ is due to the relativistic symmetry of the
crystal. Anyway since we have an ``empty" space-time with
{\it a priori\/} no relativistic crystalline structures,
the crystalline relativistic symmetry can come from 
crystalline (periodic in space-time, namely waves)
electromagnetic  or $\B$ fields dressing the particle,
{\it e.g.} dressed by the crystalline part of the external
Faraday tensor fields or the more general $\B$ fields.
Hence, this last interaction is not really due to the
electromagnetic or ``$\B$" interaction but to the need for
a local relativistic symmetry conservation.\par Another
strange situation would be $\xi^h\equiv x^h_0=\,i^h(s)$
meaning $(\chi^{k,h})$ would be a magnetic moment, and if
in addition the symmetric part of
$\B$ vanishes, then the expression
\begin{equation*}
\sum_{i,h=1}^n\,(x_0^h\,u^i-u^h\,x_0^i)\,\F_{i,h}
\end{equation*}
would describe an interaction of an electromagnetic
fields with a magnetic moment as in the Larmor precession.
Let us notice to finish that we have also to consider in
addition the Lagrangian density
\eqref{LAB} for the dynamics of $\A$ and
$\B$.\par\medskip
More generally, the Euler-Lagrange equations associated to
$S_2$ would be a system of geodesic equations  with
Riemann-Christoffel symbols associated to $\nu$ and such
that (with \(\nu^{ij}\simeq\omega^{ij}\) at first order 
and assuming the $\chi$'s being
constants)
\begin{equation}
\frac{\;\,du^r}{ds}=\frac{1}{2}\sum^n_{j,k=1}
\mathcal{P}_{jk}^r\;u^ju^k+
\sum_{i,h,s,t=1}^n\left(\xi_0\,\omega^{rs}\,\F_{i,s}+
k_0\,\xi^h\,(\delta^r_h\,\A_i-
\A_t\,\omega^{rt}\,\omega_{hi})\right)u^i\,,
\label{geod}
\end{equation}
with
\begin{equation*}
\mathcal{P}_{jk}^r=\chi^r(\partial_k\A_j+\partial_j\A_k)+
\sum_{h=1}^n\chi^{r,h}
\left(\partial_j\B_{k,h}+\partial_k\B_{j,h}\right)\,,
\end{equation*}
and $\A$ and $\B$ satisfying the first and second sets of
differential equations. Then the tensor $\mathcal{P}$
would be associated to gravitation fields, providing other physical interpretations for
the tensors $\chi$.\par Nevertheless, the equations \eqref{geod} are deduced
without the conditions \eqref{thomgen} and
\eqref{thomas}. Taking them into account would lead to a
modification of the action $S_2$ by adding Lagrange
multipliers $\lambda_0$ and $\lambda_k$ ($k=1,\cdots,n$)
in the Lagrangian density definition, and changing the
variable of integration we can define a new action:
\begin{equation*}
{S}_{\tau}^2=\int_0^{\ell}
\left\{m+\sum_{i=1}^n\epsilon_{i}(\xi_0,\xi^h)\,U^i
-\left(\lambda_0\xi_0+\sum_{k=1}^n\lambda_k\xi^k\right)
\right\}d\tau\,,
\end{equation*}
with the measure $d\tau\equiv ds/\gamma$ on the hyperboloid
$H(1,n-1)$, and
$\epsilon_{i}(\xi_0,\xi^h)\equiv\sum_{k=1}\epsilon_{ki}U^k$
with relations \eqref{prec}. The associated
Euler-Lagrange equations would be analogous to
\eqref{geod} with $U$ instead of $u$ but with additional
terms coming from the precession. Moreover, since we have the
constraint
${\parallel}U{\parallel}^2=1$, we need a new Lagrangian
multiplier denoted by $m$ ($>0$) ! That also means we do
computations on the projective spaces $H(1,n-1)$ of the
tangent spaces and make variational calculus with an
induced metric $\nu_P=\nu(P_\ast,P_\ast)$ with $P$ a
particular projector, and a measure $P^\ast(ds)=d\tau$.
From this view point, the Lagrange multipliers appear to
be non-homogeneous coordinates of these projective spaces.
Alternatively from
\eqref{lnubis} we can take the action
\begin{equation*}
S^1_\tau\equiv\int_0^{\ell}\left\{m+
\xi_0\,.
\sum_{i=1}^n\A_i\,U^i+
\xi^h\,.
\sum_{i=1}^n
U^i\,\B_{i,h}-
\left(\lambda_0\xi_0+\sum_{k=1}^n\lambda_k\xi^k\right)
\right\}d\tau\,.
\end{equation*}
Then the variational calculus
would also lead to additional precession equations giving
torsion as mentioned in the comments about the equations
\eqref{torsion}. Again the torsion is not
related to the unification but  to parallel
transports on manifolds which is a well-known geometrical
fact \cite{dieudon}.\par\medskip
\remark All functions and applications are defined at $x_0$ 
and not at $x\in U_{x_0}$, meaning we must write Taylor series at $x_0$ to deduce their
values at $x$. In other words, physical evaluations of $X\equiv x-x_0$ must be done, and
all the variables at $x$ ($\chi$, $\A$, etc\dots) would be functions of $x_0$ and $X$.
But $X$ is not physically evaluable, since we would be able to travel in time to go to $x$
and come back to $x_0$, or more generally to turn ``freely" in space-time around $x_0$.
Hence we can only do measurements at $x_0$ of light rays coming from $x$. Then we must
build up, from the light wave vectors denoted $k$, the position of $x$. It follows that we
would have to find a sheaf map from a neighborhood of $x_0$ to a neighborhood of the origin
of the tangent space
$T_{x_0}\m$ containing the $k$'s. This can be only done locally, involving  we
shall call position vectors what would be in fact vectors in the tangent space and not the
$X$'s. That is an equivalence principle. Then wave functions, for instance, would depend, at
least, on
$x_0$ and tensorial variables. In particular, this is expressed in the series $s$ for the
functions
$\alpha$ when inverting
$x$ and $x_0$.
\section{Conclusion}
In fact in this work, using the Pfaff systems theory
instead of the Kumpera-Malgrange-Spencer theory of Lie
equations, we studied the formal solutions of the
conformal Lie system with respect to the Poincar\'e one.
More precisely, we determined the difference between
these two sets of formal solutions. It is described by a
``relative" set of no Lie PDE's, namely the ``c" system.
We studied these two Lie systems of Lie equations because
of their occurrences in physics and  particularly in
electromagnetism as well as in Einsteinian relativity.
We just made the assumption that the {\it``substrat
space-times"\/} would be equivariant with
respect to the conformal pseudogroup and  setting its Riemann scalar
curvature  to be a constant $k_0$.  Then, we built up  differential complexes and tried
to give with some theorems, interpretations in physics of
the various tensors coming from the relative complex.
This was only a ``classical approach'' and quantization
didn't seem to appear. Nevertheless a deeper analysis of
the latter conformal actions, which are of Polyakov type
in dimension 4 or also a 1-acyclic cocycle, in the framework 
of the A. Connes
non-commutative geometry would work out likely a
quantization of the space of leafs of the hyperboloid
space of the 4-vectors
$U$. Alternatively another approach, with the concept of
time operator and the signature of the metric (which might
change in this model), can be consider in the framework of
Kolmogorov flows~\cite{mpc}.


\end{document}